\documentclass[12pt]{article}
\usepackage{amsfonts}
\usepackage{amsmath}
\usepackage{amssymb}
\usepackage{mathptmx}
\usepackage[dvips]{color}
\usepackage{epsfig}
\usepackage{graphicx}
\newcommand{\calA}{{\cal A}}

\newcommand{\calL}{{\cal L}}

\newcommand{\kslash}{/\!\!\!\!\!k}

\newcommand{\qslash}{/\!\!\!\!\!q}

\newcommand{\kvec}{{\bf k}}
\newcommand{\uvec}{{\bf u}}

\newcommand{\xvec}{{\bf x}}

\newcommand{\lam}{\lambda}

\newcommand{\Br}{\textrm{Br}}

\newcommand{\eV}{{\rm eV}}
\newcommand{\GeV}{{\rm GeV}}

\newcommand{\TeV}{{\rm TeV}}
\newcommand{\rmv}{{\rm v}}

\begin{document}
\baselineskip=16pt

\pagenumbering{arabic}

\vspace{1.0cm}

\begin{center}
{\Large\sf Radiative and flavor-violating transitions of leptons
from interactions with color-octet particles}
\\[10pt]
\vspace{.5 cm}

{Yi Liao$^{a,b,c}$\footnote{liaoy@nankai.edu.cn}, Ji-Yuan Liu$^b$}

{$^a$ Center for High Energy Physics, Peking University, Beijing
100871, China\\
$^b$ School of Physics, Nankai University, Tianjin 300071,
China\\
$^c$ Kavli Institute for Theoretical Physics China, CAS, Beijing
100190, China}

\vspace{2.0ex}

{\bf Abstract}

\end{center}

It has been recently proposed that neutrino mass could originate
from Yukawa interactions of leptons with new colored particles. This
raises the interesting possibility of testing mass generation
through copious production of those particles at hadron colliders. A
realistic assessment of it however should take into account how
large those interactions could be from available precision results.
In this work we make a systematic analysis to the flavor structure
in Yukawa couplings, provide a convenient parametrization to it, and
investigate the rare radiative and pure leptonic decays of the muon
and tau leptons. For general values of parameters the muon decays
set stringent constraints on the couplings, and all rare tau decays
are far below the current experimental sensitivity. However, there
is room in parameter space in which the muon decays could be
significantly suppressed by destructive interference between colored
particles without generically reducing the couplings themselves.
This is also the region of parameters that is relevant to collider
physics. We show that for this part of parameter space some tau
decays can reach or are close to the current level of precision.

\begin{flushleft}
% corresponding to: (PACS 2010 version)
% Neutrino mass and mixing
% Decays of leptons
% Neutrino interactions


\end{flushleft}

\newpage

\section{Introduction}
\label{sec:intro}

It has been verified by oscillation and other low energy experiments
that neutrinos have tiny and non-degenerate masses, yet the
mechanism for mass generation has remained mysterious. It is
generally believed that the new physics relevant to the tiny mass
lies beyond the standard model (SM) and its effects may be
systematically accounted for by high dimensional operators. Indeed,
viewing SM as an effective field theory at low energies one can
write down a unique dimension five operator in terms of the Higgs
and lepton doublet fields that generates the neutrino mass upon
spontaneous breaking of electroweak symmetries
\cite{Weinberg:1979sa}. It is interesting that there are only three
realizations of the operator at tree level \cite{Ma:1998dn} which
correspond exactly to the three types of seesaw models
\cite{type1,type2,Foot:1988aq}. In each case, a single
representation under the SM gauge group, $SU(3)_C\times
SU(2)_L\times U(1)_Y$, is prescribed for the new fields to link the
Higgs and lepton doublets.

In the above seesaw models, the tininess of the neutrino mass is
attributed to a huge scale of new physics or feeble interactions
that induce the operator. In either case it would be difficult to
detect low energy effects of new physics beyond neutrino mass. One
way to relieve the tension is to consider certain radiative origin
of neutrino mass \cite{Zee:1980ai}. By prescribing two or more types
of representations but excluding those utilized in the three seesaw
models, it is possible to generate tiny neutrino mass without
requiring all couplings to be diminishingly small or all new
particles to be inaccessibly heavy. This is even so when the mass
originates from a two-loop \cite{Zee:1985id,Babu:1988ki} or
three-loop effect \cite{Aoki:2008av}. These radiative mechanisms
usually employ small representations of the electroweak group. The
other way to generate neutrino mass without sacrificing too much in
couplings or heavy masses has been suggested recently
\cite{Babu:2009aq}. By assigning larger representations, it is
possible to forbid the dimension five operator at tree level so that
the operator relevant for neutrino mass first appears at dimension
seven.

In all of the above mechanisms it has been tacitly assumed that the
particles responsible for neutrino mass generation do not
participate strong interactions. But this could well be an easy
prejudice as there is no experimental hint for it at all. The idea
that neutrino mass may originate from interactions with colored
particles becomes especially relevant now. These particles if not
very heavy can be copiously produced at the LHC so that the origin
of neutrino mass could potentially be tested there. A concrete model
in this spirit has been recently proposed by Fileviez Perez and Wise
\cite{FileviezPerez:2009ud}. Since the colored fields must appear in
pair in interactions responsible for neutrino mass, the minimal
choice includes both scalar and fermionic degrees of freedom. To
avoid chiral anomaly from the new fermions, they considered the
simplest option that the fermions belong to the adjoint
representation of $SU(3)_C$ and are neutral under $U(1)_Y$. This in
turn singles out naturally the octet scalars that were previously
introduced in the quark sector \cite{Manohar:2006ga, Burgess:2009wm}
in the framework of minimal flavor violation
\cite{Chivukula:1987py}. The potential relevance of the model to
leptogenesis has been discussed in Ref. \cite{Losada:2009yy}.

The interactions that generate neutrino mass and mixing generically
mediate lepton flavor violating (LFV) transitions and anomalous
magnetic moments of charged leptons. While the measurements in the
$e\mu$ sector have reached an impressive level of precision
\cite{Odom:2006zz, Bennett:2006fi, Brooks:1999pu, Bellgardt:1987du},
the upper bounds on LFV decays of the tau lepton are improving
rapidly at the $B$ factories, see Refs. \cite{Aubert:2009tk,
Hayasaka:2007vc, Marchiori:2009ww, Miyazaki:2007zw} for the
precision frontier. And even more stringent limits on some of them
are expected in the near future. These constraints from low energy
processes will significantly affect the feasibility of testing
neutrino mass mechanisms at colliders. The purpose of the current
work is to investigate systematically those processes in the color
octet model suggested in Ref. \cite{FileviezPerez:2009ud}, and to
see if there is any room in the parameter space that could be
relevant to collider physics. The LFV decays have been extensively
studied in various models of neutrino mass and mixing. As a few
examples, we mention Refs. \cite{Hisano:1995cp} in supersymmetric
models, \cite{Kakizaki:2003jk, Abada:2008ea, Abada:2007ux} in seesaw
models, \cite{Bu:2008fx} in a model of mirror fermions
\cite{Hung:2006ap}, and in \cite{Choudhury:2006sq} in a little Higgs
model. We refer to the reviews \cite{Kuno:1999jp} for a more
complete list of references.

The paper is organized as follows. In the next section we introduce
the octet model and provide a convenient parametrization to the
Yukawa couplings. The radiative and pure leptonic transitions are
calculated analytically in section \ref{sec:analytic}, and the
parameter space is then discussed in section \ref{sec:numerical}. We
summarize our main results in the last section. Some phase space
integrals are evaluated in Appendix A.

\section{Parametrization of couplings in octet model}
\label{sec:model}

As we briefly reviewed in section \ref{sec:intro}, the octet model
\cite{FileviezPerez:2009ud} introduces the color-octet scalars and
fermions on top of the SM Higgs field $H$ and lepton fields,
$L_{L\alpha}$, $\ell_{R\alpha}$, where $L(R),~\alpha$ refer to
chirality and flavor respectively. The octet scalars, $S_{ar}$, have
the same quantum numbers under $SU(2)_L\times U(1)_Y$ as $H$. Here
$a$ is the color index and $r$ enumerates the scalars. The octet
fermions are neutral under $U(1)_Y$ but may be a singlet or triplet
of $SU(2)_Y$, which we denote as the two cases:
\begin{eqnarray}
\textrm{case A:}&&\rho_{ax},\nonumber\\
\textrm{case B:}&&\chi_{ax}=\left(\!\!\!\begin{array}{cc}
\frac{1}{\sqrt{2}}\chi^0_{ax}&\chi^+_{ax}\\
\chi^-_{ax}&-\frac{1}{\sqrt{2}}\chi^0_{ax}\end{array}\!\!\!\right),
\end{eqnarray}
where $x$ enumerates the fermions. We assume arbitrarily $\rho$ and
$\chi$ are right-handed. While $\rho$ and $\chi^0$ are Majorana
fields, $\chi^\pm$ being real under $SU(3)_C$ will be paired into a
Dirac field. The Yukawa couplings of leptons in SM and the
additional terms in the octet model are
\begin{eqnarray}
-\calL^\textrm{Yuk}_\textrm{SM}&=&
y_{\alpha\beta}\overline{L_{L\alpha}}H\ell_{R\beta}
+\textrm{h.c.},\nonumber\\
-\calL^\textrm{Yuk}_\textrm{A}&=&
z_{\alpha x}^r\overline{L_{L\alpha}}\tilde S_{ar}\rho_{ax}
+\textrm{h.c.},\\
-\calL^\textrm{Yuk}_\textrm{B}&=&%
z_{\alpha x}^r\overline{L_{L\alpha}}\chi_{ax}\tilde S_{ar}
+\textrm{h.c.},\nonumber
\end{eqnarray}
where $\tilde S_{ar}=i\tau_2S_{ar}^*$ and summation over repeated
indices is implied.

The charged leptons become massive when $H$ develops a vacuum
expectation value, $\rmv/\sqrt{2}$. Since neutrinos are massless at
tree level, the diagonalizing matrices for $\ell$ can be absorbed by
redefinition of the fields. We assume this has been done already.
For generally non-degenerate scalars or fermions there is no
symmetry mixing them. Since the octet fermions are neutral under
$U(1)_Y$ and either an adjoint or a singlet of $SU(2)_L$, rephasing
is not possible for them. For the scalars, we assume their phase
convention has been fixed in the potential \cite{Manohar:2006ga}.
Thus what we can do at most is to rephase the fields $\ell_L$ (and
$\ell_R$ identically), leaving the remaining magnitudes and phases
in $z$ being physical parameters.

The neutrinos may gain mass from radiative corrections. Since there
are no further degrees of freedom for them to pair with, the mass
must necessarily be of Majorana nature. For this to be possible, the
lepton number conservation has to be violated. This happens when the
Yukawa couplings are augmented by the following term in the
potential with a single octet scalar \cite{Manohar:2006ga}:
\begin{eqnarray}
-V\supset -\frac{1}{2}\lambda(S_a^\dagger H)^2+\textrm{h.c.},
\end{eqnarray}
where $\lambda$ is real by convention. It turns out that for two out
of the three neutrinos to become massive, the minimal choice is to
have either one octet scalar plus two octet fermions or the other
way around \cite{FileviezPerez:2009ud}. To avoid the complicated
mixing amongst scalars, we shall stick here to the former choice. We
shall drop from now on the index $r$ while $x$ assumes values
$1,~2$. In this minimal scenario, the $z$ couplings contain nine
physical parameters, i.e., six magnitudes plus three phases, and
will be parameterized later in this section.
\begin{figure}
\centering
\includegraphics[width=9cm]{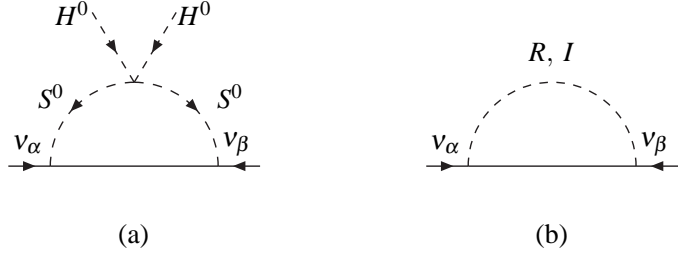}
\caption{Diagrams for neutrino masses.} %
\label{fig1}
\end{figure}

The neutrino mass matrix may be obtained by first computing the
dimension five operator, $\overline{\nu_\beta}P_L\nu_\alpha H^0H^0$,
depicted in Fig. 1(a), where the virtual fermion can be $\rho_x$
(case A) or $\chi^0_x$ (case B). The result is
\cite{FileviezPerez:2009ud},
\begin{eqnarray}
M_{\alpha\beta}^\nu=\sum_{x=1}^2z^*_{\alpha x}z^*_{\beta x}\mu_x,
\end{eqnarray}
where
\begin{eqnarray}
\mu_x=\xi C\frac{\lam\rmv^2m_x}{32\pi^2m_0^2}%
\frac{1+r_x(\ln r_x-1)}{(r_x-1)^2},~r_x=\frac{m^2_x}{m_0^2}.
\label{eq_mu_approx}
\end{eqnarray}
Here $C=N_C^2-1=8$ is a color factor, and $\xi=1~(1/2)$ in case A (
B). We have denoted the masses of $S^0$ and $\rho_x$ ($\chi^0_x$) by
$m_0$ and $m_x$ respectively. Since the mass splitting between the
neutral and charged octet particles arises at a higher order, we
shall ignore it and use the same notation for $S^\pm$ (and
$\chi^\pm_x$ in case B). The mass matrix can also be obtained by
calculating the self-energy of neutrinos at zero momentum shown in
Fig. 1(b), where the decomposition $S^0_a=(R_a+iI_a)/\sqrt{2}$ has
been made. The result is to replace the above $\mu_x$ by the
following one:
\begin{eqnarray}
\mu_x=\frac{\xi C}{32\pi^2}
\frac{m_x}{(m_x^2-m_R^2)(m_x^2-m_I^2)}\left[%
m_R^2m_x^2\ln\frac{m_x^2}{m_R^2}-m_I^2m_x^2\ln\frac{m_x^2}{m_I^2}
+m_R^2m_I^2\ln\frac{m_R^2}{m_I^2}\right],%
\label{eq_mu_exact}
\end{eqnarray}
where $m_{R,I}$ are the masses of $R,~I$, whose splitting is
measured by the $\lam$ term in the potential,
\begin{eqnarray}
m^2_R-m^2_I=\lam\rmv^2.
\end{eqnarray}
The result in eq (\ref{eq_mu_approx}) is an approximation to eq
(\ref{eq_mu_exact}) in the limit of small mass splitting,
$|\lam\rmv^2|\ll(m^2_R+m^2_I)/2$. Since $\lam$ is a measure of
lepton number violation, it should be naturally small. We shall work
in this limit below.

With two octet fermions the matrix $M^\nu$ is degenerate and has a
zero eigenvalue. This arises because it is a product of a matrix of
lower rank with its transpose. Noting $\mu_x/\lam>0$ for arbitrary
masses, we can express it as $M^\nu=ZZ^T$, where
$Z=(\uvec_1^*,\uvec_2^*)$ is a $3\times 2$ matrix in terms of the
column vectors, $\uvec_x=\sqrt{\mu_x}z_{\alpha x}$ (not summed over
$x$) for $\lam>0$, or $\uvec_x=\sqrt{-\mu_x}(iz_{\alpha x})$ for
$\lam<0$. The factors of $i$ in the latter case will not affect our
later results which involve $z$ always in the form of $z^*_{\alpha
x}z_{\beta x}$, and can thus be ignored. This implies that we can
restrict ourselves in phenomenological analysis to $\lam>0$ without
loss of generality. Diagonalization by the leptonic mixing matrix
$U_\textrm{PMNS}$ yields two possible spectra of normal or inverted
hierarchy:
\begin{eqnarray}
\textrm{NH:}&&m_{\nu_1}=0,~m_{\nu_2}=\lam_-,~m_{\nu_3}=\lam_+,
\nonumber\\
\textrm{IH:}&&m_{\nu_3}=0,~m_{\nu_1}=\lam_-,~m_{\nu_2}=\lam_+,
\end{eqnarray}
where $\lam_+>\lam_->0$ are the two non-zero eigenvalues. For both
hierarchies, we have $U_\textrm{PMNS}=U_\textrm{D}E_\nu$, where
$U_\textrm{D}$ is the analog of the CKM matrix with three angles
$(\theta_{12},\theta_{23},\theta_{13})$ and a Dirac CP phase
$\delta$, and $E_\nu=\textrm{diag}(1,e^{i\alpha},1)$ contains a
single Majorana phase. We write in terms of the column vectors:
\begin{eqnarray}
U_\textrm{PMNS}=(\xvec_1,\xvec_2,\xvec_3),
\end{eqnarray}
with $\xvec_i^\dagger\xvec_j=\delta_{ij}$. It can be readily
verified that the zero eigenvalue requires the corresponding row
(first for NH or third for IH) of $U_\textrm{PMNS}^TZ$ to vanish.
This means that $Z$ can be generally parameterized in terms of the
other two column vectors in $U_\textrm{PMNS}$:
\begin{eqnarray}
\textrm{NH:}&&\uvec_1=c_-\xvec_2+c_+\xvec_3,
~\uvec_2=d_-\xvec_2+d_+\xvec_3,\nonumber\\
\textrm{IH:}&&\uvec_1=c_-\xvec_1+c_+\xvec_2,
~\uvec_2=d_-\xvec_1+d_+\xvec_2.
\end{eqnarray}
The coefficients $c_\pm$, $d_\pm$ can be determined in terms of the
eigenvalues $\lam_\pm$ and a free complex parameter. We find that
the following parametrization is convenient and applies to both
hierarchies:
\begin{eqnarray}
&&c_-=\sqrt{\lam_-}\frac{2t}{1+t^2},~
d_-=\sqrt{\lam_-}\frac{1-t^2}{1+t^2},\nonumber\\
&&c_+=\sqrt{\lam_+}\frac{1-t^2}{1+t^2},~
d_+=\sqrt{\lam_+}\frac{-2t}{1+t^2},%
\label{eq_para}
\end{eqnarray}
where $t$ is complex. And the Yukawa couplings now become
\begin{eqnarray}
z_{\alpha x}=\frac{\uvec_{\alpha x}}{\sqrt{|\mu_x|}}.
\end{eqnarray}
The physical parameters in $z_{\alpha x}$ have been traded for two
neutrino masses $\lam_\pm$, three angles $\theta_{ij}$, one Dirac
phase $\delta$, one Majorana phase $\alpha$, and one complex number
$t$.

The diagonalization of leptons causes no other changes in gauge
interactions but attaching $U_\textrm{PMNS}$ to their charged
currents. In particular, the neutral currents of leptons are flavor
diagonal. We can therefore restrict ourselves to interactions
involving octet particles for the effects that are suppressed by SM
interactions. Introducing the four-component fields,
\begin{eqnarray}
\nu=\left(\begin{array}{c}\nu_L\\ \nu_L^C\end{array}\right),~%
\rho_a=\left(\begin{array}{c}\rho_{aR}^C\\
\rho_{aR}\end{array}\right),~%
\chi_a^0=\left(\begin{array}{c}\chi_{aR}^{0C}\\
\chi_{aR}^0\end{array}\right),~ \chi^-_a=\left(\begin{array}{c}
\chi^{+C}_{aR}\\
\chi_{aR}^-\end{array}\right),
\end{eqnarray}
and using $\bar fg=\overline{g^C}f^C$, the Yukawa couplings are
modified to ($P_R=(1+\gamma_5)/2$),
\begin{eqnarray}
-\calL^\textrm{Yuk}_\textrm{A}&=&
\overline{\nu}U^\dagger_\textrm{PMNS}z
P_R\rho_aS_a^{0*}-\overline{\ell}zP_R\rho_aS_a^-
+\textrm{h.c.},\nonumber\\
-\calL^\textrm{Yuk}_\textrm{B}&=&
\frac{1}{\sqrt{2}}\overline{\nu}U^\dagger_\textrm{PMNS}z
P_R\chi_a^0S_a^{0*}%
-\overline{\chi_a^-}(U^\dagger_\textrm{PMNS}z)^TP_R\nu S_a^-
\nonumber\\
&&+\overline{\ell}zP_R\chi_a^-S_a^{0*}%
+\frac{1}{\sqrt{2}}\overline{\ell}zP_R\chi_a^0S_a^-%
+\textrm{h.c.}.%
\label{eq_yuk}
\end{eqnarray}
Note in passing that the massless neutrino does not couple to color
octet fermions.
\begin{figure}
\centering
\includegraphics[width=14cm]{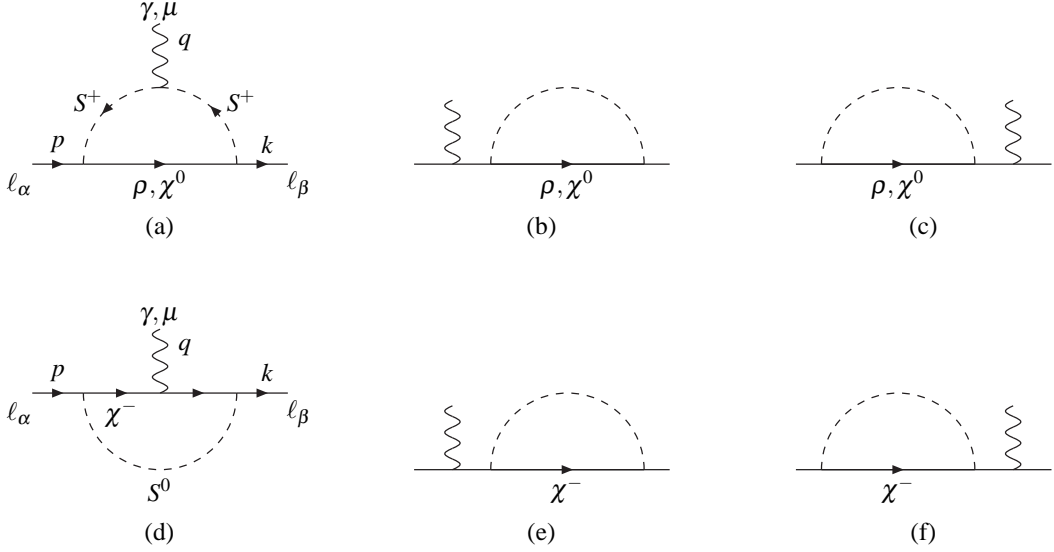}
\caption{Diagrams for radiative transitions.} %
\label{fig2}
\end{figure}

\section{Radiative and pure leptonic transitions}
\label{sec:analytic}

A direct consequence of the interactions shown in eq (\ref{eq_yuk})
is the occurrence of LFV transitions of the charged leptons,
$\ell_\alpha\to\ell_\beta\gamma$ and
$\ell_\delta\to\ell_\alpha\ell_\beta\bar\ell_\gamma$, and the
contribution to related quantities like the anomalous magnetic
moment and the muonium-anti-muonium transition rate. We start with
the transition $\ell_\alpha(p)\to\ell_\beta(k)+\gamma^{(*)}(q)$,
which also contributes to the pure leptonic decay when the photon
converts to a pair of leptons. Our sign convention for the QED
coupling is, $\calL_\textrm{QED}=-eA_\mu\bar\ell\gamma^\mu\ell$.

The Feynman diagrams are shown in Fig. 2, where (a)-(c) appear in
both cases A and B and (d)-(f) occur only for case B. Working to the
first nontrivial order in the external momenta, a straightforward
calculation yields the amplitude:
\begin{eqnarray}
-\frac{(4\pi)^2m_0^2}{Ce}\calA_\mu&=&%
z_{\beta x}z^*_{\alpha x}\Big\{\left[\xi F_1(r_x)
+2(1-\xi)G_1(r_x)\right]\bar u(k)
P_R(q^2\gamma_\mu-q_\mu\qslash)u(p)\nonumber\\
&&+\left[\xi F_2(r_x)+2(1-\xi)G_2(r_x)\right]%
\bar u(k)(m_\alpha P_R+m_\beta P_L)i\sigma_{\mu\nu}q^\nu u(p)\Big\},
\end{eqnarray}
where summation over $x$ is implied, and the loop functions are
\begin{eqnarray}
F_1(x)&=&\frac{1}{36(x-1)^4} \left[2-9x+18x^2-11x^3+6x^3\ln
x\right],\nonumber\\
F_2(x)&=&\frac{1}{12(x-1)^4}\left[1-6x+3x^2+2x^3-6x^2\ln x\right],
\nonumber\\
G_1(x)&=&\frac{1}{36(x-1)^4}\left[-16+45x-36x^2+7x^3-12\ln
x+18x\ln x\right],\nonumber\\
G_2(x)&=&\frac{1}{12(x-1)^4}\left[-2-3x+6x^2-x^3-6x\ln x\right].
\end{eqnarray}
Some comments are in order. That the above Lorentz structures
satisfy Ward identity serves as a useful check to our calculation.
The $F_{1,2}$ terms from Figs (2a)-(2c) occur for both cases A and B
while the $G_{1,2}$ terms from Figs (2d)-(2f) appear only in case B.
This is controlled by the switch parameter $\xi$ introduced earlier.
Since we have ignored the mass splitting between the neutral and
charged particles, the loop functions depend only on the mass ratios
of the fermions to the scalar. While $F_{1,2}(x),~G_1(x)>0$ and
$G_2(x)<0$, they all decrease monotonically in magnitude as $x$
increases. Since the contributions from the real and imaginary parts
of $S^0$ in Figs. (2d)-(2f) simply add, when their mass splitting is
relevant the exact result can be obtained by averaging the shown one
over their masses.

From the above result we obtain the lepton anomalous magnetic moment
and the branching ratio for the transition
$\ell_\alpha\to\ell_\beta\gamma$:
\begin{eqnarray}
a(\ell_\alpha)&=&\frac{1}{\pi}\frac{m_\alpha^2}{m_0^2}
\sum_{x=1}^2|z_{\alpha x}|^2\big[\xi
F_2(r_x)+2(1-\xi)G_2(r_x)\big],\\
\Br(\ell_\alpha\to\ell_\beta\gamma)&=&%
\Br(\ell_\alpha\to\nu_\alpha\ell_\beta\bar\nu_\beta)
\nonumber\\
&&\times\frac{12\alpha_\textrm{EM}}{\pi G_F^2m_0^4}
\Big|\sum_{x=1}^2z_{\beta x}z^*_{\alpha x} \big[\xi
F_2(r_x)+2(1-\xi)G_2(r_x)\big]\Big|^2,
\end{eqnarray}
where $\alpha_\textrm{EM}$ is the fine structure constant. We have
used
$\Gamma(\ell_\alpha\to\nu_\alpha\ell_\beta\bar\nu_\beta)
=G_F^2m_\alpha^5/(192\pi^3)$
and neglected the final state masses in phase space integration. The
decay $\mu\to e\gamma$ was also computed for case A (i.e., $\xi=1$)
in Ref. \cite{Losada:2009yy}, but our result differs from theirs.

The pure leptonic decay
$\ell_\delta(p)\to\ell_\alpha(k_1)\ell_\beta(k_2)\bar\ell_\gamma(k_3)$
receives a contribution from box diagrams, in addition to the one
from the off-shell radiative transition computed above when
$\ell_\beta=\ell_\gamma$ or $\ell_\alpha=\ell_\gamma$. The box
diagrams are shown in Fig. 3, where the arrows indicate the flow of
negative charge. Fig. 3(a) appears for both cases A and B while Fig.
3(b) and (c) occur only for case B. In the approximation of
neglecting the external momenta compared with the heavy octet
masses, Fig. 3(b) and (c) cancel each other, leaving us with the
result:
\begin{eqnarray}
\textrm{Fig. 3}=\frac{C\xi^2}{m^2_0i(4\pi)^2}
z_{\alpha x}z^*_{\delta x}z_{\beta y}z^*_{\gamma y}%
\bar u(k_1)\gamma_\mu P_Lu(p)~\bar u(k_2)\gamma^\mu P_Lv(k_3)%
H(r_x,r_y),
\end{eqnarray}
where summations over $x$ and $y$ are implied, and the loop function
is
\begin{eqnarray}
H(x,y)=\frac{1}{4(x-y)} %
\left[\frac{1}{1-x}-\frac{x\ln x}{1-x}+\frac{x\ln x}{(1-x)^2}%
-\frac{1}{1-y}+\frac{y\ln y}{1-y}-\frac{y\ln y}{(1-y)^2}\right],
\end{eqnarray}
which is symmetric and positive-definite. The result has the correct
anti-symmetry under the interchanges $\alpha\leftrightarrow\beta$
and $k_1\leftrightarrow k_2$ upon using Fierz identity.
\begin{figure}
\centering
\includegraphics[width=11cm]{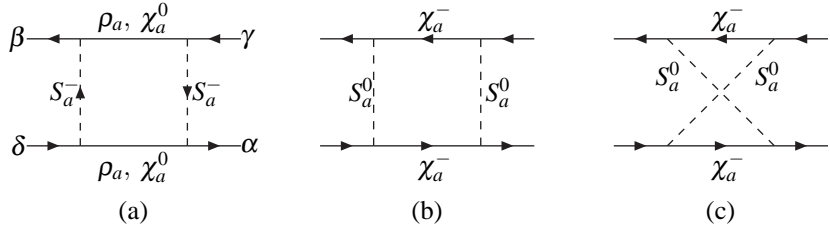}
\caption{Additional box diagrams for pure leptonic decays.} %
\label{fig3}
\end{figure}

With three generations of leptons there are three possible flavor
configurations for the final states from a fixed initial state:
\begin{eqnarray}
(1)&&\ell_\delta(p)\to\ell_\alpha(k_1)\ell_\alpha(k_2)
\bar\ell_\gamma(k_3),~\ell_\alpha\ne\ell_\gamma,\nonumber\\
(2)&&\ell_\delta(p)\to\ell_\alpha(k_1)\ell_\beta(k_2)
\bar\ell_\beta(k_3),~\ell_\alpha\ne\ell_\beta,\nonumber\\
(3)&&\ell_\delta(p)\to\ell_\alpha(k_1)\ell_\alpha(k_2)
\bar\ell_\alpha(k_3).
\end{eqnarray}
We list their decay amplitudes separately. For decay (1), only the
box diagram contributes:
\begin{eqnarray}
i\calA(1)=\frac{2C\xi^2}{m^2_0 i(4\pi)^2}z_{\alpha x}z^*_{\delta
x}z_{\alpha y}z^*_{\gamma y}\bar u_1\gamma_\mu P_Lu(p)\bar
u_2\gamma^\mu P_Lv_3H(r_x,r_y),
\end{eqnarray}
where the factor 2 takes into account the identical particles in the
final state upon using Fierz identity. For decay (2), both radiative
and box diagrams contribute. It is important to make sure that their
relative sign is correct:
\begin{eqnarray}
i\calA(2)=i\calA_\triangle+i\calA_\Box,
\end{eqnarray}
where upon using equations of motion and Fierz identity and denoting
$s_{ij}=(k_i+k_j)^2$,
\begin{eqnarray}
i\calA_\triangle&=&%
\frac{-Ce^2z_{\alpha x}z^*_{\delta x}}{m^2_0i(4\pi)^2}%
\bar u_2\gamma^\mu v_3~\bar u_1\Big[P_R\gamma_\mu\big(\xi
F_1(r_x)+2(1-\xi)G_1(r_x)\big)\nonumber\\%
&&+(m_\delta P_R+m_\alpha P_L)i\sigma_{\mu\nu}(k_2+k_3)^\nu
s_{23}^{-1}
\big(\xi F_2(r_x)+2(1-\xi)G_2(r_x)\big)\Big]u(p),\nonumber\\
i\calA_\Box&=&%
\frac{C\xi^2}{m^2_0i(4\pi)^2}z^*_{\beta y}z^*_{\delta x}
\left(z_{\alpha x}z_{\beta y}+z_{\beta x}z_{\alpha y}\right) \bar
u_1\gamma_\mu P_Lu(p)~\bar u_2\gamma^\mu P_Lv_3H(r_x,r_y).%
\label{eq_decay2}
\end{eqnarray}
Finally, the amplitude for decay (3) can be obtained from eq
(\ref{eq_decay2}) by substitutions and anti-symmetrization:
\begin{eqnarray}
\calA(3)=\left.\calA(2)\right|_{\beta=\alpha}-(k_1\leftrightarrow
k_2).\label{eq_A3}
\end{eqnarray}

The branching ratio for the process (1) can be worked out
straightforwardly using the phase space integral $I_3$ defined and
calculated in Appendix A,
\begin{eqnarray}
\Br(1)=\Br(\ell_\delta\to\ell_\alpha\nu_\delta\bar\nu_\alpha)
\frac{C^2\xi^4}{2^{10}\pi^4m_0^4G_F^2} \Big|\sum_{x,y}z_{\alpha
x}z^*_{\delta x}z_{\alpha y}z^*_{\gamma y} H(r_x,r_y)\Big|^2.
\end{eqnarray}
The decay rate for the process (2) is more complicated, both because
it has radiative and box contributions and because there is a
logarithmic singularity in the mass ($m_\beta$) of leptons connected
to the virtual photon. Care must be exercised in dropping $m_\beta$
in order not to miss terms at the considered order. We find it
convenient to decompose the amplitude in the form:
\begin{eqnarray}
C^{-1}m^2_0(4\pi)^2\calA(2)&=&
\Big(B+T_1-T_2m^2_\delta s^{-1}_{23}\Big)%
\bar u_2\gamma^\mu P_Lv_3\bar u_1\gamma_\mu P_Lu(p)
\nonumber\\
&&+\Big(T_1-T_2m^2_\delta s^{-1}_{23}\Big)\bar u_2\gamma^\mu
P_Rv_3\bar u_1\gamma_\mu P_Lu(p)\nonumber\\
&&+2T_2m_\delta s^{-1}_{23}\bar u_2\kslash_1v_3\bar u_1P_Ru(p),
\end{eqnarray}
where
\begin{eqnarray}
B&=&-\xi^2z^*_{\beta y}z^*_{\delta x}\left(z_{\alpha x}z_{\beta
y}+z_{\beta x}z_{\alpha y}\right)H(r_x,r_y),
\nonumber\\
T_1&=&e^2z_{\alpha x}z^*_{\delta x} \big(\xi
F_1(r_x)+2(1-\xi)G_1(r_x)\big),
\nonumber\\
T_2&=&e^2z_{\alpha x}z^*_{\delta x} \big(\xi
F_2(r_x)+2(1-\xi)G_2(r_x)\big).%
\label{eq_BT}
\end{eqnarray}
The decay rate is, in terms of the integrals in Appendix A,
\begin{eqnarray}
\frac{\Gamma(2)}{m_\delta}\frac{m^4_0(4\pi)^4}{C^2m_\delta^4}&=&
|B+T_1|^24I_3+|T_1|^24I_3-\textrm{Re}(BT_2^*)4I_2
-\textrm{Re}(T_1T_2^*)4I_1\nonumber\\
&&+|T_2|^2\left[2J_1-8J_2+4K\right],%
\label{eq_rate2}
\end{eqnarray}
and the branching ratio is
\begin{eqnarray}
\Br(2)&=&\Br(\ell_\delta\to\ell_\alpha\nu_\delta\bar\nu_\alpha)
\frac{C^2}{2^{11}\pi^4m_0^4G_F^2}\bigg\{ |B+T_1|^2+|T_1|^2
-4\textrm{Re}(BT_2^*)\nonumber\\
&&-8\textrm{Re}(T_1T_2^*)+\Big[-\frac{14}{3}
+8\ln\frac{m_\delta^2}{4m_\beta^2}\Big]|T_2|^2 \bigg\}.
\end{eqnarray}

Although the amplitude for the process (3) contains twice as many
terms as for the process (2), its decay rate is not much more
difficult to calculate. Each of the two terms in eq (\ref{eq_A3})
when squared separately contributes the same to the rate, i.e., same
as shown in eq (\ref{eq_rate2}) upon setting $\beta=\alpha$ and
including a factor $\frac{1}{2}$ for identical particles. Although
their interference has both $s_{23}^{-1}$ and $s_{13}^{-1}$ terms,
the potential singularities do not overlap in kinematics. We can
therefore do algebra with $m_\alpha=0$ without losing singular terms
of the form $\ln m_\alpha^2$. Using the symmetry of phase space we
find all those terms are cancelled in the interference, yielding a
regular contribution to the rate. The sum is
\begin{eqnarray}
\frac{\Gamma(3)}{m_\delta}\frac{m^4_0(4\pi)^4}{C^2m_\delta^4} &=&
|B+T_1|^28I_3+|T_1|^24I_3-\textrm{Re}(BT_2^*)8I_2
-\textrm{Re}(T_1T_2^*)4(I_1+I_2)\nonumber\\
&&+|T_2|^2\Big[2I_2+2J_1-8J_2+4K\Big],%
\label{eq_rate3}
\end{eqnarray}
where $B$, $T_1$, and $T_2$ are obtained from eq (\ref{eq_BT}) by
setting $\beta=\alpha$. The branching ratio is,
\begin{eqnarray}
\Br(3)&=&\Br(\ell_\delta\to\ell_\alpha\nu_\delta\bar\nu_\alpha)
\frac{C^2}{2^{11}\pi^4m_0^4G_F^2}\bigg\{
2|B+T_1|^2+|T_1|^2\nonumber\\
&&-8\textrm{Re}(BT_2^*)-12\textrm{Re}(T_1T_2^*)
+\Big[-\frac{8}{3}+8\ln\frac{m_\delta^2}{4m_\alpha^2}
\Big]|T_2|^2\bigg\}.
\end{eqnarray}
The decay rate for $\mu\to 3e$ from similar Lorentz structures was
also calculated long ago in Ref \cite{Cheng:1977nv}. Their result
coincides with ours only upon including the $K$ term as shown in eq
(\ref{eq_rate3}). This term was easily missed since it is apparently
of a higher order in $m_\alpha^2$.

Finally, the process (1) also implies an effective interaction that
can induce the muonium-anti-muonium oscillation:
\begin{eqnarray}
\calL_\textrm{eff}=-G_{M\bar M}\bar\mu\gamma^\alpha P_Le
~\bar\mu\gamma_\alpha P_Le,
\end{eqnarray}
where the effective Fermi constant is,
\begin{eqnarray}
\frac{G_{M\bar M}}{\sqrt{2}}=\frac{C\xi^2}{m^2_0(4\pi)^2}
\sum_{x,y=1}^2z_{\mu x}z^*_{e x}z_{\mu y}z^*_{e y}H(r_x,r_y).
\end{eqnarray}

\section{Numerical analysis}
\label{sec:numerical}

Before we embark on numerical discussion, we make some general
remarks on the results obtained so far.  This will help us identify
potentially interesting regions in parameter space. The starting
point is the induced neutrino masses that set a basic constraint on
the free parameters in the octet model. Since the neutrinos are
extremely much lighter than the octet particles, the effects due to
Yukawa couplings may be relevant only when the $\lam$ parameter is
tiny. With fixed neutrino and octet masses, we have $|z|\propto(\lam
C)^{-1/2}$, where $C$ is the dimension of the real representation to
which the new colored particles belong. Therefore, a smaller $\lam$
tends to enhance the box diagrams more than the radiative ones,
while a larger representation tends to suppress the former while
leaving the latter intact. But in practice the box diagrams never
dominate leptonic transitions for perturbative $z$ parameters when
the stringent constraints from muon decays are taken into account.
As a matter of fact, we find that those constraints are so strong
that the branching ratios of all rare tau decays are generally much
below the level to be accessible in the near future even when some
$z$ parameters have a magnitude of order one. This arises from the
fact that the lepton mixing is very close to the tri-bimaximal
pattern whose entries are either order one or zero. This feature
generally carries over to the $z$ parameters though the latter
involve other parameters. For the tri-bimaximal mixing, we have for
NH
\begin{eqnarray}
z_{\alpha 1}^\textrm{tb}
=\frac{t}{1+t^2}\sqrt{\frac{\lam_-}{\mu_1}}\left(
\begin{array}{l}
+\frac{2}{\sqrt{3}}e^{i\alpha}\\
+\frac{2}{\sqrt{3}}e^{i\alpha}+\frac{\eta}{\sqrt{2}}r_\nu\\
-\frac{2}{\sqrt{3}}e^{i\alpha}+\frac{\eta}{\sqrt{2}}r_\nu
\end{array}\right),~
z_{\alpha 2}^\textrm{tb}
=\frac{t}{1+t^2}\sqrt{\frac{\lam_-}{\mu_2}}\left(
\begin{array}{l}
+\frac{\eta}{\sqrt{3}}e^{i\alpha}\\
+\frac{\eta}{\sqrt{3}}e^{i\alpha}-\frac{2}{\sqrt{2}}r_\nu\\
-\frac{\eta}{\sqrt{3}}e^{i\alpha}-\frac{2}{\sqrt{2}}r_\nu
\end{array}\right),
\end{eqnarray}
and for IH
\begin{eqnarray}
z_{\alpha 1}^\textrm{tb}=\frac{t}{1+t^2}
\sqrt{\frac{\lam_-}{\mu_1}}\left(
\begin{array}{l}
+\frac{\eta}{\sqrt{3}}r_\nu e^{i\alpha}+\frac{2\sqrt{2}}{\sqrt{3}}
\\
+\frac{\eta}{\sqrt{3}}r_\nu e^{i\alpha}-\frac{\sqrt{2}}{\sqrt{3}}
\\
-\frac{\eta}{\sqrt{3}}r_\nu e^{i\alpha}+\frac{\sqrt{2}}{\sqrt{3}}
\end{array}\right),~
z_{\alpha 2}^\textrm{tb}
=\frac{t}{1+t^2}\sqrt{\frac{\lam_-}{\mu_2}}\left(
\begin{array}{l}
-\frac{2}{\sqrt{3}}r_\nu e^{i\alpha}+\frac{\sqrt{2}\eta}{\sqrt{3}}
\\
-\frac{2}{\sqrt{3}}r_\nu e^{i\alpha}-\frac{\eta}{\sqrt{6}}\\
+\frac{2}{\sqrt{3}}r_\nu e^{i\alpha}+\frac{\eta}{\sqrt{6}}
\end{array}\right),
\end{eqnarray}
where $r_\nu=\sqrt{\lam_+/\lam_-}>1$ by definition and
$\eta=(1-t^2)/t$ is complex.

Since the constraints from the decays $\mu\to e\gamma,~3e$ make the
majority of parameter space practically inaccessible, we turn to
consider the possibility that the dominant contributions to those
processes may be cancelled between the two octet fermions. This will
then impose a relation among various parameters. Since the
tri-bimaximal pattern serves as an excellent approximation to the
mixing matrix and simplifies the analysis considerably, we shall
determine the regions in parameter space where the cancellation
occurs for the pattern. This will be employed later as a guide to
scan parameters in interesting intervals where some processes might
be accessible without breaking the stringent bounds on the muon
decays. In addition, we find that the heavy scalar limit $m_0\gg
m_{1,2}$ is particularly interesting. In the limit the radiative
decays are independent of the scalar mass while the pure box induced
decays are power enhanced. This apparently strange behavior is of
course due to the fixed neutrino masses. But we should also keep in
mind that we cannot take the limit literally for numerical analysis
since $z$ being proportional to $m_0$ would exceed the perturbative
regime.

In the heavy octet scalar limit, $r_x\ll 1$, all loop functions are
independent of $r_x$ except that $G_1(r_x)$ has a residual $\ln r_x$
dependence. The suppression of muon decays then simplifies to the
$(\mu e)$ cancellation condition:
\begin{eqnarray}
z_{\mu 1}^{\textrm{tb}*}z_{e1}^\textrm{tb}+z_{\mu
2}^{\textrm{tb}*}z_{e2}^\textrm{tb}=0,%
\label{eq_cancel}
\end{eqnarray}
which involves the parameters $\alpha$, $r_\nu$, $\eta$, and
$R=m_1/m_2$. In case A all leading terms are cancelled, while in
case B there remain $\ln r_x$ terms from $G_1$ which are significant
only when $m_{1,2}$ are well separated. Our later numerical
discussion will not involve this situation. Eq (\ref{eq_cancel}) can
be solved exactly. Consider the NH first. The existence of solutions
to eq (\ref{eq_cancel}) requires that
\begin{eqnarray}
\cos(2\alpha)\le f(R),~
f(x)=\frac{1}{2x}\bigg[\frac{3r_\nu^2}{8x}(1-x^2)^2-(1+x^2)\bigg].
\end{eqnarray}
For $R\not\in(R_-,R_+)$, the above is automatically fulfilled while
for $R\in(R_-,R_+)$ it has to be checked. Here we have denoted
$R_\pm=\Big[\sqrt{1+2/(3r_\nu^2)}\pm\sqrt{2/(3r_\nu^2)}\Big]^2$,
which are roughly $0.51$ and $1.97$ using the best-fit values for
neutrino masses \cite{Maltoni:2004ei}, $\Delta
m^2_\textrm{sol}=7.6\times 10^{-5}~\eV^2$, $|\Delta
m^2_\textrm{atm}|=2.4\times 10^{-3}~\eV^2$. Denoting
$\eta=|\eta|e^{i\beta}$, the phase $\beta$ is determined uniquely
for given $(\alpha,R)$ by
\begin{eqnarray}
\tan\beta=\frac{1-R}{1+R}\tan\alpha,~
c_1=\sqrt{6}r_\nu(1-R)\frac{\cos\beta}{\cos\alpha}<0,
\end{eqnarray}
and the real positive solutions for $|\eta|$ are
\begin{eqnarray}
|\eta|_\pm=\frac{1}{2R}\Big[-c_1\pm\sqrt{c_1^2-16R}\Big].
\end{eqnarray}
To each $\eta$ correspond two values of $t$, $t_\pm=\frac{1}{2}
\Big[-\eta\pm\sqrt{\eta^2+4}\Big]$. There are thus generally four
solutions to the $(\mu e)$ cancellation condition in eq
(\ref{eq_cancel}), which will be named $t_{1,2}$ (from $|\eta|_+$)
and $t_{3,4}$ (from $|\eta|_-$). Note that $t_1t_2=t_3t_4=-1$. The
relation is not accidental but reflects a symmetry in our
parametrization in eq (\ref{eq_para}): when $t$ is replaced by
$-1/t$ our $z$ only flips its sign.

The IH case is solved similarly. The existence of solutions to eq
(\ref{eq_cancel}) requires that
\begin{eqnarray}
\cos(2\alpha)\le g(R),~g(x)=\frac{1}{2x}\bigg[
\frac{r_\nu^2(1-x^2)^2}{8(x-r_\nu^2)(1-r_\nu^2x)}-(x^2+1)\bigg].
\end{eqnarray}
Defining $R_\pm=R_0\pm\sqrt{R_0^2-1}$ with
$R_0=(4r_\nu^4+r_\nu^2+4)/(9r^2_\nu)$ and noting
$r_\nu^2>R_+>R_->r_\nu^{-2}$, the requirement is automatically
fulfilled for $R\not\in(R_-,R_+)$ while it has to be checked for
$R\in(R_-,R_+)$. The latter interval is very narrow in IH since
$R_-\approx 0.985$ and $R_+\approx 1.015$. There are generally two
solutions to $\eta$ (except at $R=r_\nu^2$),
\begin{eqnarray}
\eta_\pm=\frac{1}{2}e^{i\beta}\big[-c_1\pm\sqrt{c_1^2-4c_0}\big],
\end{eqnarray}
where
\begin{eqnarray}
&&\beta=\arctan\bigg(\frac{R-1}{R+1}\tan\alpha\bigg)
\in(-\pi/2,\pi/2),\nonumber\\
&&c_1=\frac{\sqrt{2}r_\nu(1-R)}{(r_\nu^2-R)}\frac{\cos\beta}{\cos\alpha},
~c_0=\frac{4(r_\nu^2R-1)}{(r_\nu^2-R)}.
\end{eqnarray}
The solutions for $t$ are also denoted by $t_{1,2,3,4}$ as in NH
case.

We show in Fig. 4 the curves for the cancellation condition in eq
(\ref{eq_cancel}) for both NH and IH cases. In the upper panels, the
real $t$ parameter varies as a function of $R$ at $\alpha=0$. All of
four solutions are shown ($t_{1(2)}$: upper (lower) solid curve,
$t_{3(4)}$: upper (lower) dotted). Since $|f(x)|<1$ for
$x\in(R_-,R_+)$ in the NH case, there are no solutions to $t$ when
$R$ lies in the interval. The situation is similar in the IH case
though the interval becomes very narrow due to $r_\nu\approx 1$ and
$t$ varies rapidly close to the ends of the interval. When $R$ is
removed away from $R_\pm\sim 1$, it is a good approximation to
assume $r_\nu=1$, in which limit the four solutions become flat in
$R$ and saturate the values, $t=(-1\pm\sqrt{3})/\sqrt{2}$,
$\sqrt{2}\pm\sqrt{3}$. In other words, at these values of $t$ the
cancellation occurs independently of $R$ as long as $R$ is not too
close to unity. The lower panels show how the complex $t$ varies as
the Majorana phase $\alpha$ moves from $0$ to $\pi$ at $R=3$. For
clarity of illustration we only show $t_1$ (solid) and $t_3$
(dotted) for the NH case while $t_{2,4}$ can be recovered from
$-1/t_{1,3}$; similarly, for the IH case only $t_{1,4}$ (solid and
dotted) are depicted while the major portions of $t_{2,3}$ lie
outside of the displayed regions of $t$. The curves are
discontinuous at $\alpha=\frac{1}{2}\pi$, jumping from one segment
of a curve from below and at $\alpha=\frac{1}{2}\pi$ (indicated in
the figure by a dot) to the other above the value (indicated by a
cross).
\begin{figure}
\centering
\includegraphics[width=14cm]{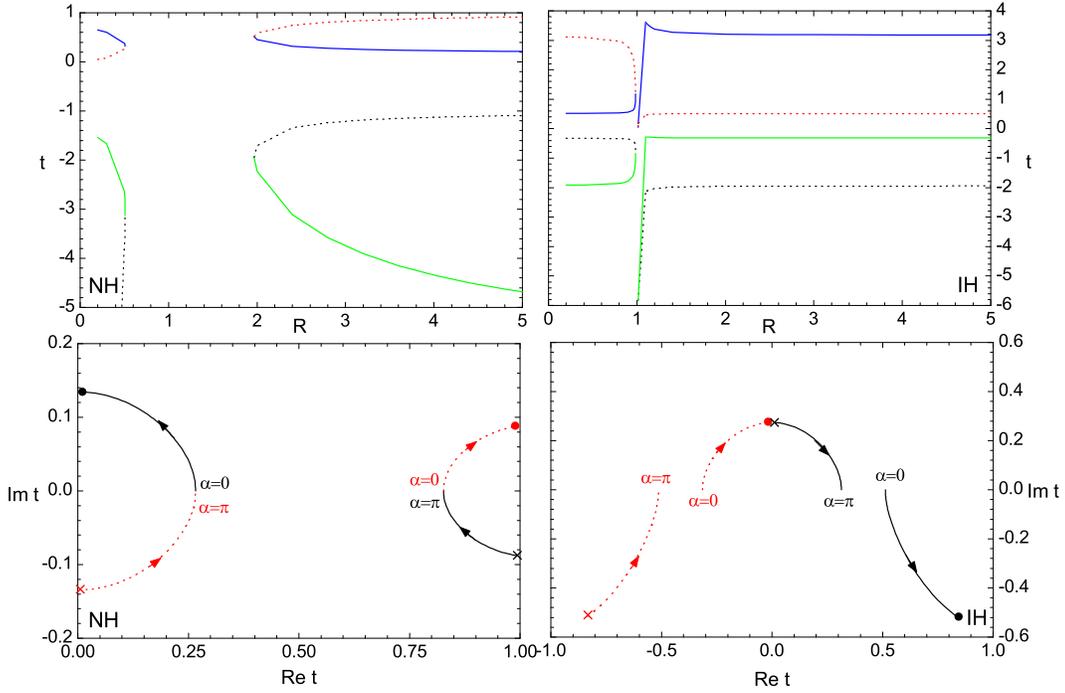}
\caption{Curves for cancellation condition in eq (\ref{eq_cancel})
are shown for both NH and IH cases. Upper panels: real $t$ varies as
a function of $R$ at $\alpha=0$. All four solutions are shown. Lower
panels: real and imaginary parts of $t$ vary as $\alpha$ changes
from $0$ to $\pi$ at $R=3$. Only two out of four solutions are
depicted. The starting and ending values of $\alpha$ are indicated.
Note the discontinuity at $\alpha=\frac{1}{2}\pi$.}
\label{fig4}
\end{figure}

Keeping in mind the cancellation curves we now display some results
on the radiative and leptonic decays. For definiteness, the octet
masses are always chosen to be
\begin{eqnarray}
m_0=10~\TeV,~m_1=400~\GeV,~m_2=200~\GeV,
\end{eqnarray}
while the $\lam$ parameter is chosen such that the strict bounds on
$\mu\to e\gamma,~3e$ decays still leave some space in which tau
decays are potentially accessible. We show in the upper panels of
Fig. 5 the branching ratios for $\mu\to e\gamma,~3e$ as a function
of $|t|$ at three points of $(\alpha,\arg t)=(0,0)$ (solid),
$(\pi/3,0.107)$ (dashed), and $(\pi/3,0.485)$ (dotted) for the NH
and tri-bimaximal mixing in case A. We have assumed $\lam=1.2\times
10^{-9}$. Although the $t$ parameter is an independent complex
parameter, we have chosen its argument properly for each value of
$\alpha$ so that $|t|$ in the interval $[0,1]$ contains one or two
points saturating the cancellation condition. At $R=2$ and
$\alpha=0$, for instance, the cancellation occurs at $t_1=0.449$,
$t_3=0.585$ plus the other two outside the interval, while at
$\alpha=\pi/3$ it occurs at $t_1=0.212\exp(0.485i)$,
$t_3=0.828\exp(0.107i)$ and two others. The current upper bounds on
the decays \cite{Brooks:1999pu, Bellgardt:1987du} are also
indicated. It is clear that the allowed ranges of $|t|$ depend
significantly on both $\alpha$ and $\arg t$. In the lower panels we
show using the same parameters the branching ratios for $\tau\to
e\mu,~3\mu$ that are closest to the current upper bounds
\cite{Aubert:2009tk, Marchiori:2009ww}. Although their variations in
$t,~\alpha$ are not as strong as designed for the muon decays, they
are still significant. At $(\alpha,\arg t)=(0,0)$ and
$(\pi/3,0.107)$ the branching ratio for $\tau\to 3\mu$ can approach
the bound in the allowed ranges of $|t|$. The results for other tau
decays and anomalous magnetic moments are about two or more orders
of magnitude smaller than their current bounds and are less
sensitive to the parameters. Fig. 6 displays similar curves for IH
at $\lam=2.5\times 10^{-9}$. The two curves in each panel correspond
to $(\alpha,\arg t)=(0,0)$ (solid) and $(\pi/3,-0.258)$ (dashed).
The muon decays are significantly suppressed around $|t|=0.51$ and
$0.61$ respectively. Their sharper spikes can also be understood
from the flatness of the $t-R$ cancellation curves in Fig. 3. In the
allowed ranges of parameters the decay $\tau\to 3\mu$ is a few times
smaller than its upper bound in the most favorable situation, while
other decays are largely unobservable.
\begin{figure}
\centering
\includegraphics[width=14cm]{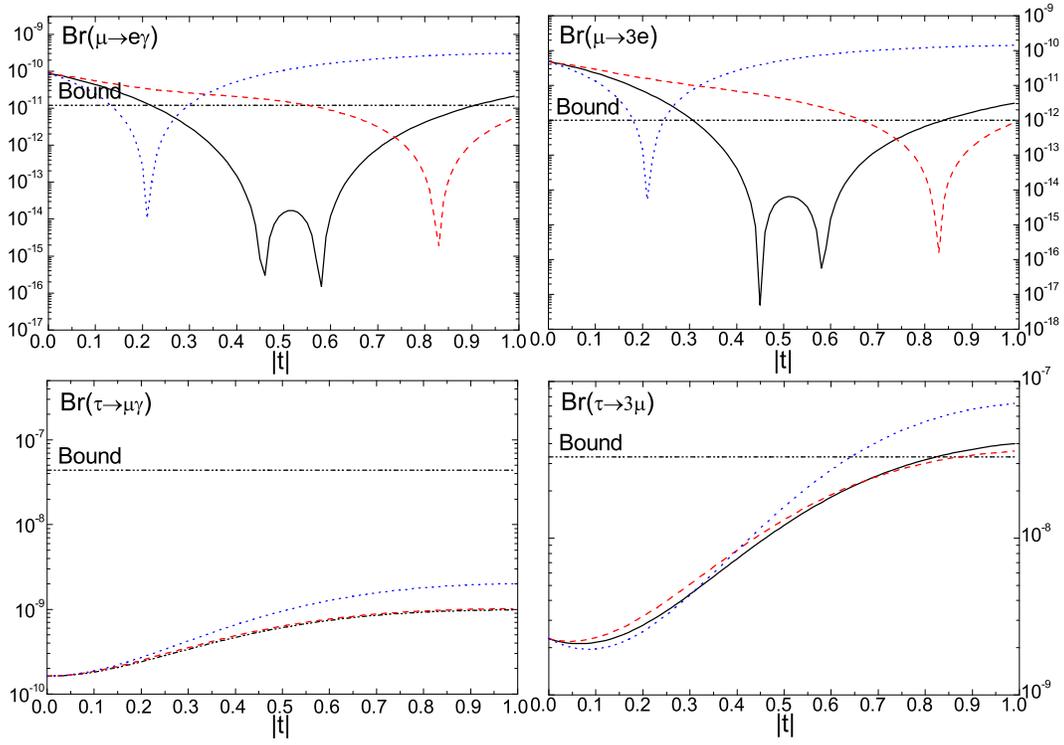}
\caption{Branching ratios as a function of $|t|$ in NH,
tri-bimaximal mixing, and case A.}%
\label{fig5}
\end{figure}

The above results are presented for the simplified scenario of
tri-bimaximal mixing. But global fittings of neutrino data generally
prefer a slight deviation from it. We investigate now how rare
decays could potentially be sensitive to the deviation. For
definiteness, we fix $\theta_{12}$ and $\theta_{23}$ to the central
values obtained in Ref. \cite{Maltoni:2004ei},
$\sin^2\theta_{12}=0.32$, $\sin^2\theta_{23}=0.50$, and vary the
small angle $\theta_{13}$ below its upper bound,
$\sin^2\theta_{13}\le 0.05$. As we pointed out earlier, for
arbitrarily chosen values of the new parameters $\alpha,~t$ and
masses $m_{0,1,2}$, all tau decays are generically too small to be
observable when the bounds on rare muon decays are respected.
Therefore, in the dominant portion of parameter space the decays
cannot be sensitive to a small parameter like $\theta_{13}$.
However, if the parameters happen to be located in the neighborhood
of cancellation curves determined by eq (\ref{eq_cancel}) and
exemplified in Fig. \ref{fig4}, the muon decays could be sensitive
to $\theta_{13}$. We show in Fig. \ref{fig7} how their branching
ratios vary as a function of $\theta_{13}$ at $\lam=10^{-8}$ and for
case B. We have set $t=0.449$ (solid line) and $0.585$ (dotted) for
NH and $t=0.511$ for IH. These values of $t$ correspond to the
points at which significant cancellation takes place in case A for
the tri-bimaximal mixing. It is interesting that $\Br(\mu\to 3e)$ in
NH reaches its minimum not at $\theta_{13}=0$ but around
$\theta_{13}\sim 0.025$. For such a `large' value of $\lam$ other
decays are simply not observable.
\begin{figure}
\centering
\includegraphics[width=14cm]{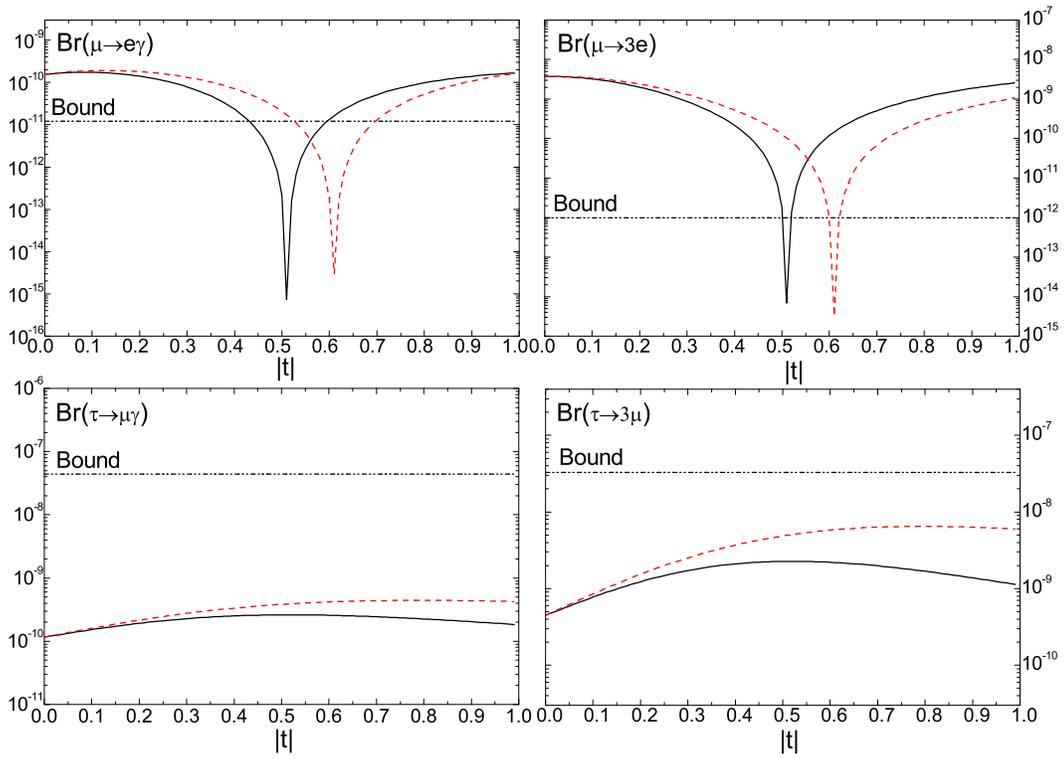}
\caption{Branching ratios as a function of $|t|$ in IH,
tri-bimaximal mixing, and case A.}
\label{fig6}
\end{figure}
\begin{figure}
\centering
\includegraphics[width=14cm]{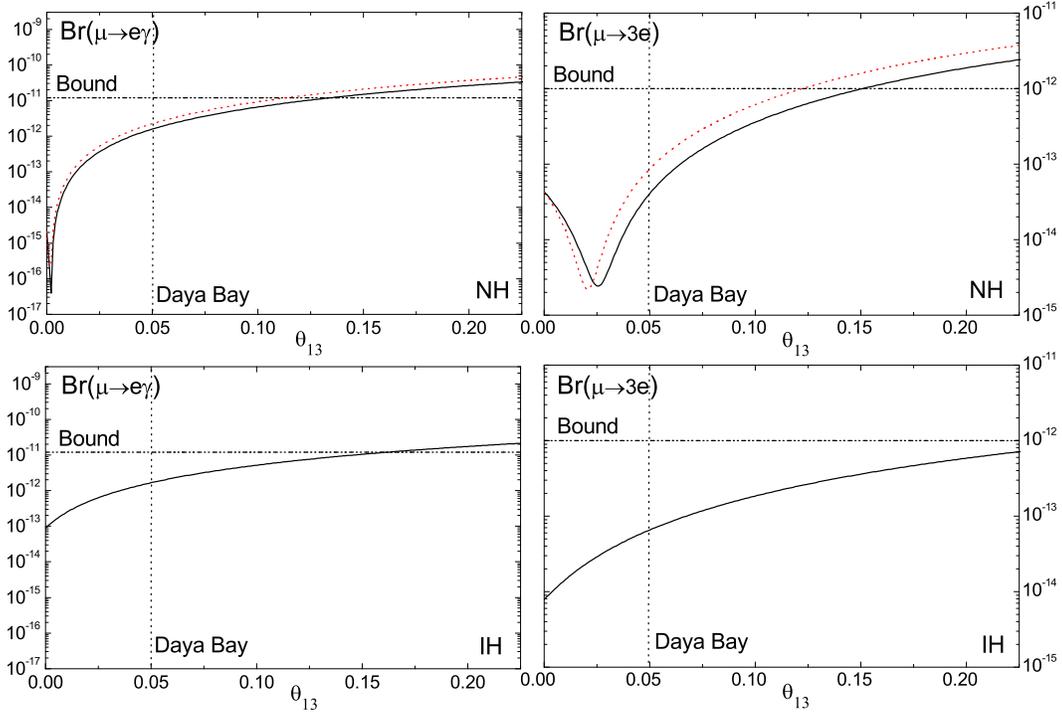}
\caption{Branching ratios in case B as a function of $\theta_{13}$
at $\sin^2\theta_{12}=0.32,~\sin^2\theta_{23}=0.50,~\delta=0$ and
$\alpha=0$. Upper panels for NH and lower ones for IH. The proposed
limit to be reachable in Daya Bay experiment is also indicated.}
\label{fig7}
\end{figure}

\section{Conclusion}

A mechanism that generates tiny neutrino mass and large lepton
mixing generically induces lepton flavor violating transitions in
the charged lepton sector, if there is any observable effect at all
of the mechanism at low energies. The transitions are extremely
suppressed by tiny neutrino masses if they are induced only by
standard gauge interactions. Their observability would thus
necessitate the existence of new particles that could interact with
neutrinos and charged leptons separately. A model of this sort has
been devised recently \cite{FileviezPerez:2009ud}. Its main merit is
that the particles responsible for the interactions are colored, so
that they could be copiously produced at hadron colliders. However,
to make contact with the origin of neutrino mass it would be
necessary to observe lepton signals produced from those colored
particles. A realistic estimate of the signals should take into
account the constraints that are already available. In this work we
have made a complete analysis of the lepton flavor structure,
provided a convenient parametrization to it, and studied
systematically the radiative and pure leptonic transitions of the
muon and tau leptons. We found that the current bounds on rare muon
decays set a stringent constraint on rare tau decays that are
generically far below the current experimental sensitivity for the
majority of parameter space. However, there still exists the
possibility that the new particles interfere destructively in the
$(e\mu)$ sector such that the muon decays are significantly
suppressed while their interactions with leptons are not necessarily
reduced. This is indeed also the region in parameter space relevant
to collider physics. We have explored this possibility for rare tau
decays and showed that the decays $\tau\to\mu\gamma,~3\mu$ are much
enhanced compared to the general case. For the normal hierarchy of
neutrino masses, the branching ratio for $\tau\to 3\mu$ could reach
the current level of precision while for the inverted hierarchy it
is a few times smaller. Considering the rapid progress made in rare
tau decays it is worthwhile pursuing further effects of this
parameter region.

%\newpage
\vspace{0.5cm}
\noindent %
{\bf Acknowledgement}

This work is supported in part by the grants NCET-06-0211,
NSFC-10775074 and NSFC-10975078.
\\

{\Large\bf Appendix A: Phase space integrals}
\\

We discuss and calculate some phase space integrals for the
three-body decay $\ell_\delta(p)\to\ell_\alpha(k_1)\ell_\beta(k_2)
\bar\ell_{\gamma}(k_3)$. Its amplitude $\calA$ originates from box
diagrams and also from radiative transitions when a pair of $\ell$
and $\bar\ell$ is connected to the virtual photon, i.e., when
$\alpha=\gamma$, or $\beta=\gamma$, or $\alpha=\beta=\gamma$. For
our purpose of estimating LFV branching ratios, it is a good
approximation to treat the final state leptons as massless. But one
must be careful with the terms related to radiative transitions when
doing algebra in $|\calA|^2$. Although the logarithmic mass
singularity, $\ln(m^2_\delta/m^2_\gamma)$, can be readily isolated,
it is easy to miss some regular terms by setting masses (of leptons
connected to the photon) to zero too early.

The spin-summed and -averaged decay rate is
\begin{eqnarray}
\Gamma&=&\frac{1}{1+\delta_{\alpha\beta}}
\frac{1}{2m_\delta}\textrm{PS}_3\sum\overline{|\calA|^2},
\end{eqnarray}
where
\begin{eqnarray}
\textrm{PS}_3&=&\int\frac{d^3\kvec_1}{(2\pi)^32E_1}
\int\frac{d^3\kvec_2}{(2\pi)^32E_2}
\int\frac{d^3\kvec_3}{(2\pi)^32E_3}(2\pi)^4\delta^4(p-k_1-k_2-k_3).
\end{eqnarray}
Using the kinematic symmetry in the final state, all required
integrals can be converted to the following ones:
\begin{eqnarray}
(p^2)^{(2,2,3)}I_{(1,2,3)}&=&\textrm{PS}_3
(p\cdot k_1,k_{23},k_1\cdot pk_{23}),\nonumber\\
(p^2)^{(1,2)}J_{(1,2)}&=&\textrm{PS}_3\frac{(k_1\cdot
p,k_{12}k_{31})}{s_{23}},\nonumber\\
p^2K&=&\textrm{PS}_3\frac{m_2^2p\cdot k_1}{s_{23}^2},
\end{eqnarray}
where $k_{ij}=k_i\cdot k_j$ and $s_{23}=(k_2+k_3)^2$. The integrals
$J_{1,2},~K$ appear only in radiative dipole transition terms, i.e.,
those proportional to $|T_2|^2$. Our convention here is that the
virtual photon is connected to the leptons $\ell_\beta(k_2)$ and
$\bar\ell_\gamma(k_3)$ with $\beta=\gamma$.

The basic integrals are,
\begin{eqnarray}
8\pi Q^2\int\frac{d^3\kvec_2}{(2\pi)^32E_2}
\int\frac{d^3\kvec_3}{(2\pi)^32E_3}(2\pi)^4\delta^4(Q-k_2-k_3)
&=&\sqrt{\lam(Q^2,k_2^2,k_3^2)},\nonumber\\
96\pi\int\frac{d^3\kvec_2}{(2\pi)^32E_2}
\int\frac{d^3\kvec_3}{(2\pi)^32E_3}
(2\pi)^4\delta^4(Q-k_2-k_3)k_2^\alpha k_3^\beta
&=&g^{\alpha\beta}Q^2+2Q^\alpha Q^\beta,
\end{eqnarray}
with $\lam(a,b,c)=a^2+b^2+c^2-2ab-2bc-2ca$. To integrate over
$\kvec_1$, we determine the interval of its magnitude to be given by
\begin{eqnarray}
|\kvec_1|=\frac{1}{2}xm_\delta,~x\in[0,x_0],~
x_0=1-4\eta_2,~\eta_2=\frac{m_2^2}{m_\delta^2}.
\end{eqnarray}
Note that it is safe to set $m_1=0$ in computing the integrals
$J_{1,2},~K$ even for the process (3). This is because the
singularities in $s_{23}$ and $s_{31}$ do not overlap, as we pointed
out in the main text. The final results are
\begin{eqnarray}
(4\pi)^3I_1&=&\frac{1}{12}+O(\eta_2),\nonumber\\
(4\pi)^3I_2&=&\frac{1}{24}+O(\eta_2),\nonumber\\
(4\pi)^3I_3&=&\frac{1}{96}+O(\eta_2),\nonumber\\
(4\pi)^3J_1&=&\frac{1}{4}\left[-\frac{3}{2}
+\ln\frac{1}{4\eta_2}\right]
+O(\eta_2),\nonumber\\
(4\pi)^3J_2&=&\frac{1}{48}
\left[-\frac{11}{6}+\ln\frac{1}{4\eta_2}\right]
+O(\eta_2),\nonumber\\
(4\pi)^3K&=&\frac{1}{16}+O(\eta_2).
\end{eqnarray}

%\vspace{0.5cm}
\noindent %
%\newpage
%\baselineskip=20pt

\end{document}